\documentclass[useAMS]{mn2e}
\usepackage{psfig}

\newif\ifAMStwofonts

\title{X-ray behaviour of Circinus X-1 - I: X-ray Dips as a diagnostic of periodic behaviour}
\author[Clarkson, Charles, Onyett]
       {W.I.~Clarkson$^{1}$\thanks{{\it Email:}
       w.i.clarkson@open.ac.uk}\thanks{{\it Current Address}: Department of
       Physics and Astronomy, The Open University, Milton Keynes MK7
       6AA, UK}, P. A.~Charles$^{1}$, N.Onyett$^{1,2}$ \\ 1. School of
       Physics and Astronomy, Southampton University, SO17 1BJ, UK \\
       2. Astronomy Centre, CPES, University of Sussex, Falmer,
       Brighton BN1 9QJ, UK}

\date{Accepted 2003 October 15.
      Received ;
      in original form 
      }

\pagerange{\pageref{firstpage}--\pageref{lastpage}}
\pubyear{2003}

\begin{document}

\maketitle

\label{firstpage}

\begin{abstract}

We examine the periodic nature of detailed structure (particularly
dips) in the RXTE/ASM lightcurve of Circinus X-1. The significant
phase wandering of the X-ray maxima suggests their identification with
the response on a viscous timescale of the accretion disk to
perturbation. We find that the X-ray dips provide a more accurate
system clock than the maxima, and thus use these as indicators of the
times of periastron passage. We fit a quadratic ephemeris to these
dips, and find its predictive power for the X-ray lightcurve to be
superior to ephemerides based on the radio flares and the full
archival X-ray lightcurve. Under the hypothesis that the dips are
tracers of the mass transfer rate from the donor, we use their
occurrence rate as a function of orbital phase to explore the (as yet
unconstrained) nature of the donor. The high $\dot{P}$ term in the
ephemeris provides another piece of evidence that Cir X-1 is in a
state of dynamical evolution, and thus is a very young post-supernova
system. We further suggest that the radio ``synchrotron nebula''
immediately surrounding Cir X-1 is in fact the remnant of the event
that created the compact object, and discuss briefly the evidence for
and against such an interpretation.

\end{abstract}


\section{Introduction}
\label{}

Cir X-1 is a galactic X-ray binary located in the galactic plane at a
distance of 6-8 kpc (Goss \& Mebold 1977). Its most well-known feature
is strong X-ray variability on a (presumed orbital) cycle of 16.6 days
(Kaluzienski et al 1976). The source distance and X-ray lightcurve
imply variation of X ray luminosity $L_x$ of a factor $\sim$100
throughout this cycle (e.g. Kaluzienski et al 1976). Radio flares were
found at peak levels of 1Jy, on a similar period to the X-ray
variations (Haynes et al, 1978), and a similar cycle is evident in the
IR activity (Glass 1994). An arcsec-scale asymmetric jet has been
imaged (Fender et al, 1998), raising the possibility that outflow from
the system is relativistic. During an interval of low X-ray activity,
EXOSAT detected eleven X-ray bursts, and their identification with Cir
X-1 strongly suggests that the compact object is a neutron star
(Tennant et al 1986a,b). Quasi-periodic oscillations (QPO's) have been
detected and characterised on two occasions (Tennant 1987, Shirey et
al 1998). The earlier EXOSAT observations suggested Atoll-like
behaviour on the X-ray colour-colour diagram, whereas RXTE/PCA
observations in the late 1990's show Z-source behaviour (Shirey, Bradt
\& Levine 1999), implying an external magnetic field strength $\ga
10^9G$ (van der Klis 1995). X-ray maxima are often preceded by dips in
the X-ray lightcurve. Spectral fits with ASCA, RXTE and BeppoSAX
during these dips are consistent with a partial covering model, in
which the X-ray emisison consists of a bright component undergoing
varying absorption, plus a fainter component not attenuated by
absorbing matter (Brandt et al 1996, Shirey, Levine \& Bradt 1999,
Iaria et al 2001).

Very little is known about the mass-losing star in this system, due to
a high but uncertain amount of extinction to the source. Early
classification of this system as an HMXB was based on an optical
identification (Whelan et al. 1977) which was later disproven (Moneti
1992) as the single object originally identified was subsequently
resolved into three distinct stars. The best optical spectra to date
are likely dominated by the accretion disk, with the donor an
insignificant component (Johnston, Fender \& Wu 1999). The highly
uncertain extinction (5 $<$ $A_v$ $<$ 11 mag) prevents the IR colours
from constraining the nature of the donor (Glass 1994). Recent
high-resolution IR spectra with IRIS2 on the AAT (Clark et al 2003)
show strong hydrogen recombination lines and low-excitation metal
lines, which resemble the spectrum of a mid-B supergiant. However the
variability of these features makes them entirely consistent also with
emission from the accretion disk and/or accretion-driven outflow (note
however that A0538-66, a Be binary with a highly eccentric 16.6 day
orbit, shows extreme variability in spectral sub-class over its
orbital cycle as a result of X-ray irradiation and reprocessing;
Charles et al 1983).

Indirect methods of determining the mass of the donor are also
indeterminate. Modeling the donor mass and eccentricity from the
optical spectra (e.g. Johnston, Fender \& Wu 1999) assumes
identification of spectral components with the donor, which is clearly
uncertain. The apparent radio association of Cir X-1 with the
supernova remnant (SNR) G 321.9-0-3 suggested an extremely high
transverse velocity of $\sim$ 450 km $s^{-1}$ due to an asymmetric
supernova kick (Stewart et al 1993). Subsequent exploration of binary
parameter space with this kick velocity as a constraint suggested both
an extremely high eccentricity ($0.90 < e < 0.94$) and low donor mass
($M_2$ likely below 1 $M_{\odot}$; Tauris et al 1999). However, a
recent HST study (Mignani et al 2002) finds no measurable proper
motion, implying an upper limit on the transverse space velocity of
$\sim$ 200 km s$^{-1}$.  This rules out the association with SNR G
321.9-0-3, thereby leaving $M_2$ and $e$ undetermined.

The Chandra X-ray Observatory (CXO) detected a remarkable set of
strong X-ray P Cygni profiles through periastron passage,
corresponding to highly ionised states of Ne, Mg, Si, S and Fe (Brandt
\& Schulz 2000). The breadth of the lines ($\sim 200-1900$ km
s$^{-1}$) and their time variability suggest identification of these
lines with highly ionised outflow from the inner region of the
accretion disk (Schulz \& Brandt 2002). Taken with the radio
observations of an expanding jet from the source (Stewart et al 1993,
Fender 2003 priv. comm), the emerging picture of Cir X-1 is of an
extremely unusual neutron star analogue to the galactic black hole
candidate microquasars such as SS433 (see Mirabel \& Rodriguez 1999
for a review of stellar jet sources).

The 16.6 day cycle has long been fit with a quadratic ephemeris, based
on radio flares measured at HARTRAO between 1978 and 1988 (Stewart et
al 1991). This ephemeris shows a comparatively high quadratic
correction, which implies a remarkably short characteristic timescale
for the period evolution $(P/2\dot{P})$ of $\sim$ 5600 years. The
longterm X-ray lightcurve of Cir X-1 over the three decades since its
identification (Margon et al 1971) was recently examined (Saz
Parkinson et al. 2003, hereafter SP03), using Fourier techniques to
determine the first ephemeris for this system based on X-ray data
alone. Period determination with this method is dominated by the X-ray
maxima (section 4), whose large intrinsic phase scatter leads to
systematic errors in the resulting ephemeris. Furthermore the average
mass transfer rate has clearly varied by factors of at least 10 on a
timescale of decades (SP03), the likely effect of which on the binary
orbit leads us to doubt the validity of fitting a single ephemeris to
all X-ray data based on the maxima alone.

In this paper we examine in detail the RXTE/ASM lightcurve of Cir X-1,
and find the periastron X-ray dips provide a superior system clock to
the X-ray maxima. Possible explanations for the phase wandering of the
X-ray maxima are discussed. We use the resulting measurement of the
intrinsic phase profile of the dips to extract information about the
orbital eccentricity $e$ and discuss the implication of the X-ray
results for the age of Cir X-1 and its future determination. In
particular, the high $\dot{P}$ term in the ephemeris suggests the
system may be in an early state of post-SN evolution (see also
SP03).

\section{RXTE/ASM Light Curve}

Launched in December 1995, the Rossi X-Ray Timing Explorer (RXTE)
carries an All Sky Monitor (ASM), which gives continuous monitoring of
the entire sky (Levine et al 1996, and references therein). Roughly
$\sim$10 readings - called ``dwells'' - are taken of a list of sources
per day, lasting about 90 seconds per dwell. Timing information is
obtained to within 1/8s, as well as crude spectral information. Data
were selected by background and quality of coverage; all points with
background level above 10 $cs^{-1}$ were rejected as part of a
filtering technnique modeled after Levine et al (2000). The ASM is
sensitive to photon energies between 1.3 and 12.1 keV, broken into
three energy channels (1.3-3.0 keV, 3.0-5.0 keV and 5.0-12.1 keV). For
the purposes of this work, we define the ``spectral hardness'' as the
ratio of activity in the upper two channels to the lower channel,
i.e. (3.0-12.1 keV) / (1.3-3.0) keV.



\begin{figure}
\begin{center}

                \psfig{file=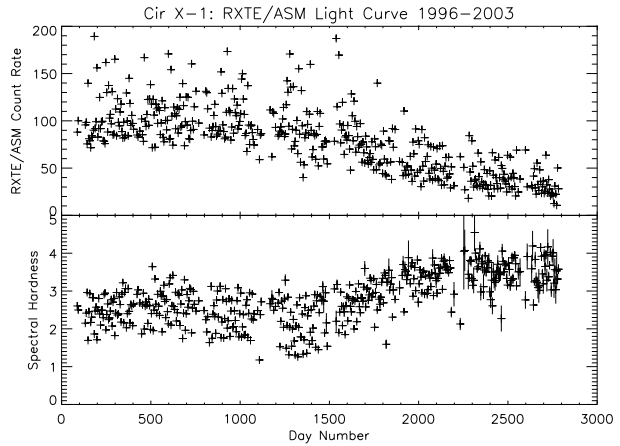,width=8.5cm}
\label{fig:out_dip}
\caption{
RXTE/ASM dataset used in this paper, binned at 5 days per
bin. Throughout the figures in this paper time is shown in units of
days elapsed since MJD 50000, i.e. 10th October 1995. {\bf Top:}
RXTE/ASM count rate. {\bf Bottom:} RXTE/ASM spectral hardness (see
Section 2).}
\end{center}
\end{figure}

\subsection{Methods of analysis}

To illustrate the phase variation of features in the lightcurve, and
how this can be used as a tool for testing an ephemeris under the
hypothesis of steady behaviour, we used the {\it dynamic
lightcurve}. This breaks the dataset into windows of sufficient length
to give good statistics, then folds these windows on the ephemeris
specified to show the evolution of phase-dependent behaviour as a
function of time. One orbital cycle per data window is sufficient to
give good signal to noise to follow the light curve structural
changes. Unless otherwise stated, the intensity of the grayscale is
directly proportional to the RXTE/ASM count rate.

We examine the usefulness of the pure RXTE/ASM lightcurve for period
determination by using a dynamic power spectrum approach (see Clarkson
et al 2003a \& b for extensive discussion of this technique). The
Lomb-Scargle periodogram code (Scargle 1982, 1989) is used in
conjunction with a sliding `data window' to produce power density
spectra for a series of overlapping stretches of the time
series. Adjustable parameters in the analysis are the length of the
data window and the amount of time by which the window is shifted in
order to obtain the overlapping stretches of data. The code accounts
for variations in the number of datapoints per interval such that the
power spectrum resolution is identical for each interval. To ensure
the 16.6 day cycle is well sampled we set our window length at 400
days. From suitable Monte Carlo simulations (see e.g. Clarkson et al
2003a) we determine a 3$\sigma$ accuracy for period determination of
$\sim$ 0.01d from sampling alone. However, it should be noted that the
orbital period of Cir X-1 is undergoing comparatively rapid evolution,
placing a limitation on the appropriateness of Fourier techniques for
long datasets.

\begin{figure}
                \psfig{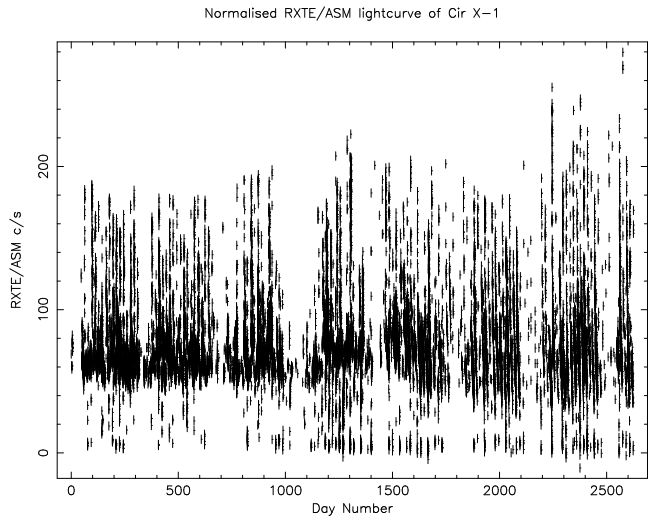}
\label{fig:fig2}
\caption{
RXTE/ASM lightcurve of Cir X-1, normalised to a running average
evaluated over intervals of 200 days (section 3.1). Count rate errors,
not overplotted in this figure for clarity, are typically 0.3
cs$^{-1}$ for the ``high'' state ranging to $\sim$ 3 cs$^{-1}$ for the
``low'' state.}
\end{figure}

Beyond Fourier techniques, much of the period determination in this
paper is performed under the hypothesis that a certain type of
behaviour should (i) be stable over time and (ii) be minimally
scattered in phase at the true period. The best-fit ephemeris is then
defined as that which minimises the phase scatter of these events over
the entire lightcurve. To determine this ephemeris, the start time,
period and quadratic correction are varied and for each combination of
parameters the resulting phase scatter is determined. The period
search was performed over a range of parameters much wider than the
variation in ephemerides to avoid bias towards any set of values. The
range of values used for the start time $MJD_0$, period $P$ and
quadratic correction, $C \simeq
\frac{1}{2}P\dot{P}$, respectively, was $\pm$ 5 d, $\pm$ 0.2d and
$\pm$ 10$^{-3}$ d. Solutions were then re-examined in finer detail,
iterating to the accuracy of the ephemeris determination. When fitting
a quadratic ephemeris, at least three orders of magnitude can be saved
from the number of computations required by noting that $P$ and $C$
can be determined independently of $MJD_0$ by subtracting this offset
with each fit so that the average phase of the lightcurve features is
always zero. Once the optimal combination of $P$ and $C$ has been
found, the best value of the offset $MJD_0$ can then be determined.

\begin{figure}
		\psfig{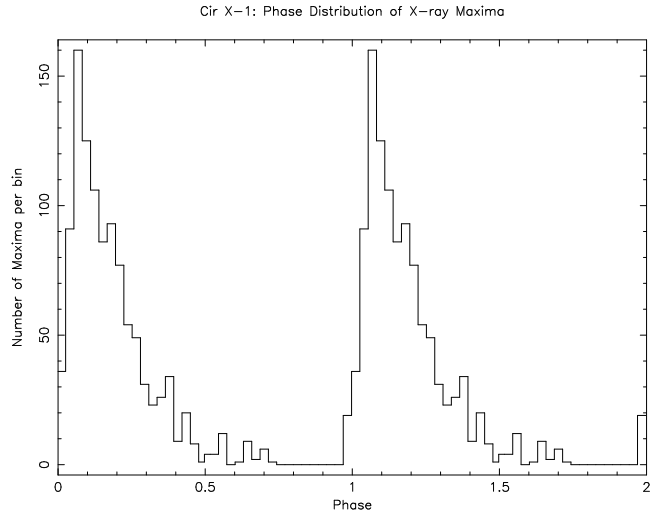}
\caption{
Distribution in phase of the X-ray maxima, folded on the radio
ephemeris (Stewart et al 1991).  }
\end{figure}

\section{X-ray behaviour}

The X-ray lightcurve can broadly be divided into three separate
behaviour types: quasi-regular dips, near phase zero of the radio
ephemeris; a ``steady'' average level beneath which the output rarely
drops, except during the dips; and periods of peak activity, during
which the output varies erratically from (1-3)$\times$ the steady
level. Furthermore, this ``steady'' level actually evolves in the
long-term lightcurve (see Figure 1) and so, for convenience in this
paper, we divide the lightcurve into three ``states'' depending on the
steady value. We shall refer to day numbers 0-700 as the ``high''
state of the source, day numbers 1800-present as the ``low'' state,
and the interval inbetween as the ``transitional'' region. We urge the
reader not to confuse these denominations with the low/hard and
high/soft states often used when describing black hole microquasars
(e.g. Tanaka \& Lewin 1995).

\begin{figure}
\begin{center}
                \psfig{file=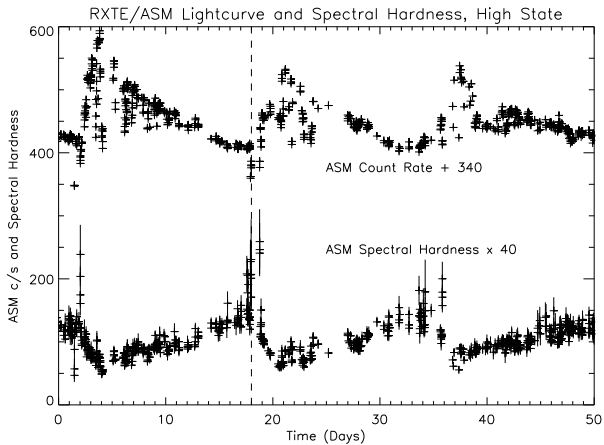,width=8.5cm}
\label{fig:out_dip}
\caption{
Top: RXTE/ASM spectral hardness for a 50d stretch of the lightcurve
starting at day number 490, during the ``high'' state. Bottom:
RXTE/ASM lightcurve during the same state, illustrating the
anticorrelation. For one cycle, phase zero according to the radio
ephemeris is plotted as a vertical dotted line.}
\end{center}
\end{figure}

\begin{figure}
\begin{center}
                \psfig{file=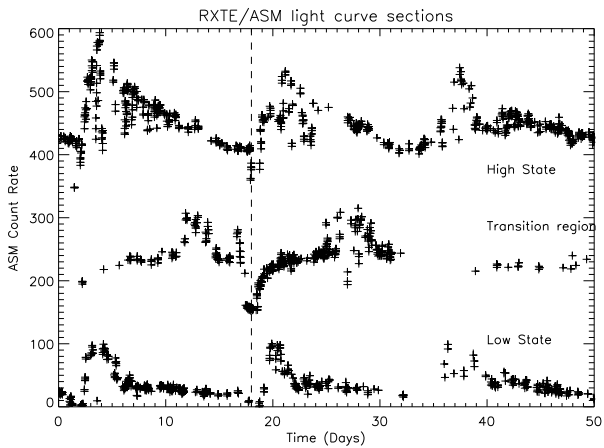,width=8.5cm}
\label{fig:fig2b}
\caption{50-day sections from the RXTE/ASM lightcurve of Cir X-1 corresponding
to the three ``states'' of the lightcurve. Top: starting at day number
490, count rate offset 340 $cs^{-1}$. Middle: starting day 1425,
offset 150$cs^{-1}$. Bottom: starting day 2350, no offset. For one
cycle, phase zero according to the radio ephemeris is plotted as a
vertical dotted line.}
\end{center}
\end{figure}

\subsection{Longterm Decline}

The steady level has decreased dramatically over the 7 years of the
RXTE/ASM dataset (Figure 1). During the ``high'' state this level
corresponds to at least 0.7 L$_{Edd}$ for a 1.4 M$_{\odot}$ neutron
star (assuming the lower limit 6 kpc distance), but currently
represents perhaps only 0.3 L$_{Edd}$. As pointed out in SP03, the
decline represents an apparent return to the X-ray levels of some
thirty years previously. This decrease is accompanied by a
corresponding increase in spectral hardness. We attempt to normalise
the longterm lightcurve by the ``steady'' level, by calculating the
average count rate over bins of 200 days, boxcar smoothing the result
and dividing the instantaneous lightcurve by this running average
lightcurve. We notice immediately that to within $\sim$ 20\% the
maximum count rate achieved at outburst scales with the running
average level (Figure 2).

\subsection{X-ray Maxima}

The X-ray behaviour at peak activity is extremely erratic and
variable, and to some extent appears dependent on the source
state. During the ``high'' state the X-ray output varies erratically
between the steady level and an envelope described by the histogram in
Figure 3. Also present during a few cycles is a smaller secondary
maximum, which appears to be similar to a secondary maximum in the
radio light curve during this state (Fender 1998). During the ``high''
state the spectral hardness is anticorrelated with the X-ray activity
(see Figure 4). The ``transitional'' interval shows similar evolution
of the spectral hardness to the ``high'' state. The X-ray maxima do
not appear to be accurate system clocks, as they occur across a large
range of phases in the 16.6 day cycle (Figure 5 and section
4). Furthermore, the shifting during this interval of the occurrence
of the maxima to phase 0.5 and later has uncovered an apparent gradual
recovery in X-ray output from a low level at phase zero (Figure 5,
middle). The lightcurve during the ``low'' state appears less erratic
than the previous two states, with fewer maxima and less phase
wandering (Figure 5, bottom).


\begin{figure}
                 \psfig{file=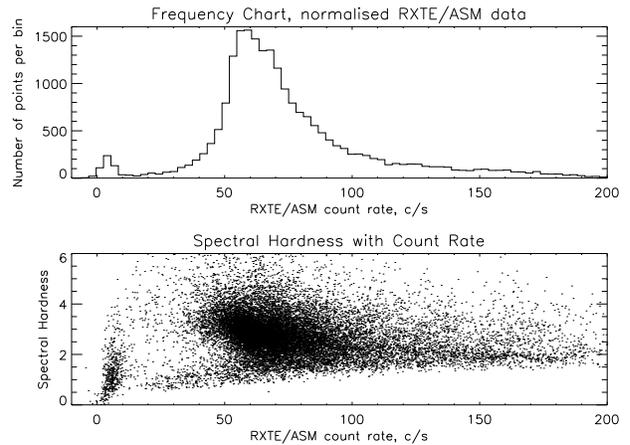,width=8.5cm}
\caption{
Top: Frequency chart for the entire RXTE/ASM dataset indicates the
existence of a separate population of low-count rate ``dips.'' Bottom:
ASM hardness-count rate plot for the RXTE/ASM dataset.}
\end{figure}

\subsection{X-ray Dips}

Phase zero in the radio ephemeris approximately predicts the
quasi-regular dips in X-ray intensity (Shirey 1998). These dips form a
separate population of datapoints, as can be seen from the
frequency-count rate chart of the ASM dataset (Figure 6). When folded
on the radio ephemeris, we see that the dips show far less scatter
about phase zero than do the maxima (Figures 7 \& 8). X-ray dip
spectra are best fit with a two-component model, with a strongly
absorbed bright component and an unabsorbed faint component (Section
1). Similar behaviour is seen during X-ray dips in RXTE/PCA
observations of black hole candidates GRO J1655-40 and 4U1630-47
(Kuulkers et al 1998).


\begin{figure}
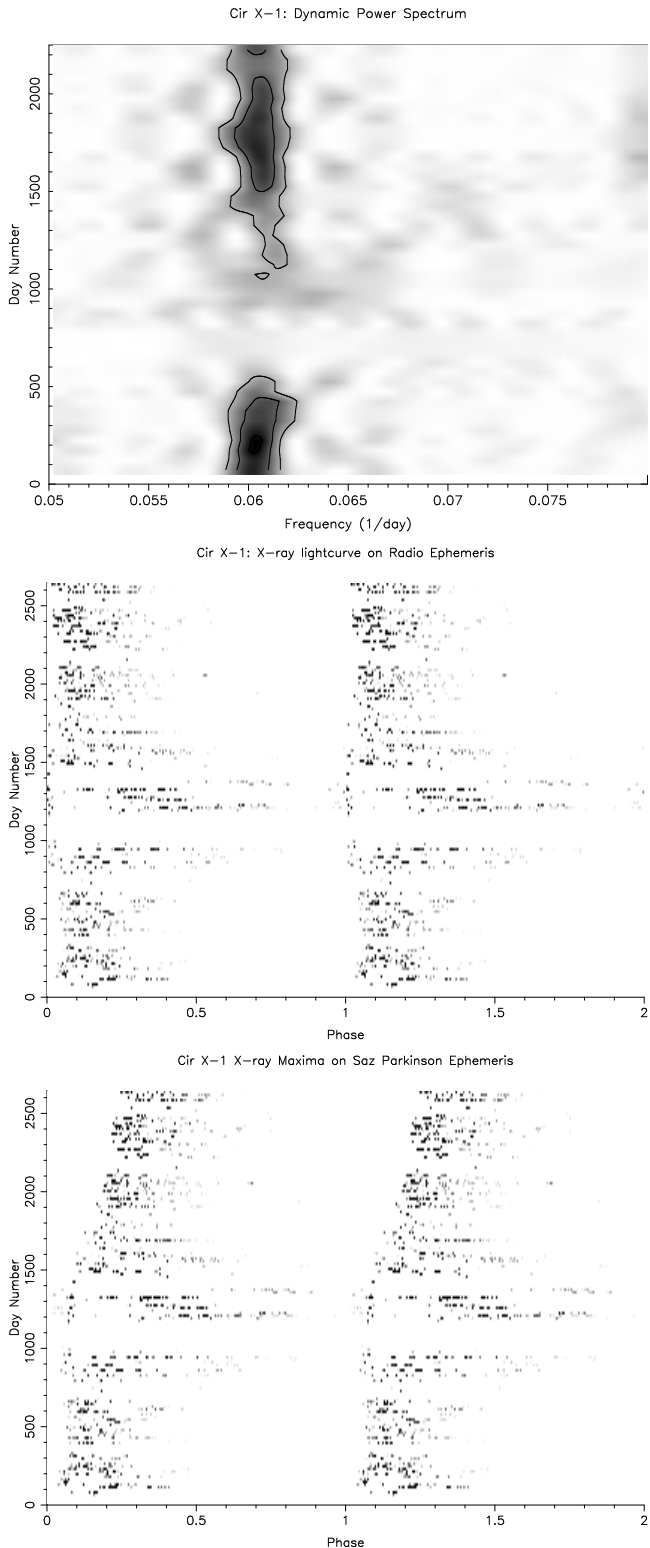

\begin{center}

                \psfig{file=fig7a.ps,width=8.5cm,angle=-90}
	\vspace{5pt}
             \psfig{file=fig7b.ps,width=8.5 cm,angle=-90}
	\vspace{5pt}

	 \psfig{file=fig7c.ps,width=8.5cm,angle=-90}
\label{fig:out_dip}
\caption{{\it Top:} Dynamic Power spectrum of the full RXTE/ASM dataset. {\it Middle:} Dynamic Lightcurve showing the X-ray maxima, folded on the radio ephemeris. {\it Bottom:} Dynamic lightcurve showing X-ray maxima, folded on the SP03 ephemeris. Note that for the dynamic power spectrum the day numbers represent the beginning of each 400-day data window in the power spectrum.}
\end{center}
\end{figure}


\section{Ephemerides}

As the best X-ray system clocks, the dips can be used to test the
various published expressions for the ephemerides of Cir X-1. Until
very recently, the most commonly used was the radio ephemeris (Stewart
et al 1991), which, when extrapolated to January 1996 (the beginning
of the RXTE/ASM dataset) can be expressed as (Shirey 1998)

\begin{equation}
	MJD_N = 50082.04+(16.54694-3.53\times10^{-5}N)N
\end{equation}

\noindent 
where $N$ is the cycle number. This ephemeris shows an r.m.s. scatter
of 0.06 days (Section 1 and Stewart et al 1991), or a 3$\sigma$
scatter in flare occurence of $\sim$0.36 days when extrapolated to the
beginning of the RXTE era. As this ephemeris is based on radio data
now more than 15 years old, its deteriorating predictive power makes
an updated ephemeris desirable. The recent FFT-based analysis of all
long-term X-ray datasets of Cir X-1 (SP03) provided the first
ephemeris based on X-ray activity alone. This ephemeris retains
$MJD_0$ from the radio ephemeris, but other than this constant offset
the two are independent. When precessed to the beginning of the
RXTE/ASM dataset, this X-ray ephemeris is expressed as:

\begin{equation}
	MJD_N = 50082.04+(16.5389-1.354\times10^{-4}N)N
\end{equation}

\noindent with 3$\sigma$ errors on the period $P$ and quadratic correction $C$ of $\sim$
1.7 $\times$ 10$^{-2}$d and $\sim$ 1.39 $\times$ 10$^{-6}$d respectively. 

We test the suitability of the x-ray maxima for period determination
by plotting the dynamic power spectrum of the entire RXTE/ASM
lightcurve of Cir X-1 with the dynamic lightcurve (Figure 7). Error in
the period determinations is manifested as a systematic drift in phase
from cycle to cycle. The transitional period, during which the maxima
wander considerably in phase, corresponds to the interval noted by
SP03, during which the 16.6d cycle is suppressed. This suggests that
the outbursts dominate the period detection from such methods, which
is why the transitional region was excised from the RXTE/ASM
lightcurve by SP03 when determining their ephemeris.

The radio ephemeris (Figures 7 \& 8) shows less of a drift,
particularly in the dynamic lightcurve of the X-ray maxima, but a
systematic drift in the occurrence of the dips is still visible from
day number 1000 onwards. The extrapolated scatter in the radio
ephemeris (section 4) leads to 3$\sigma$ phase error of $\sim
\pm$0.025 by the end of the RXTE dataset, but this is still small
compared to the systematic shift in phase of $\sim 0.1$ that we
observe. Consequently, we do not believe that this shift can be
explained by the extrapolated uncertainty in the radio ephemeris
alone. The radio ephemeris is currently quite accurate in predicting
the onset of dip events, but is expected to become less accurate as
the events drift farther backwards in phase.

\begin{figure}
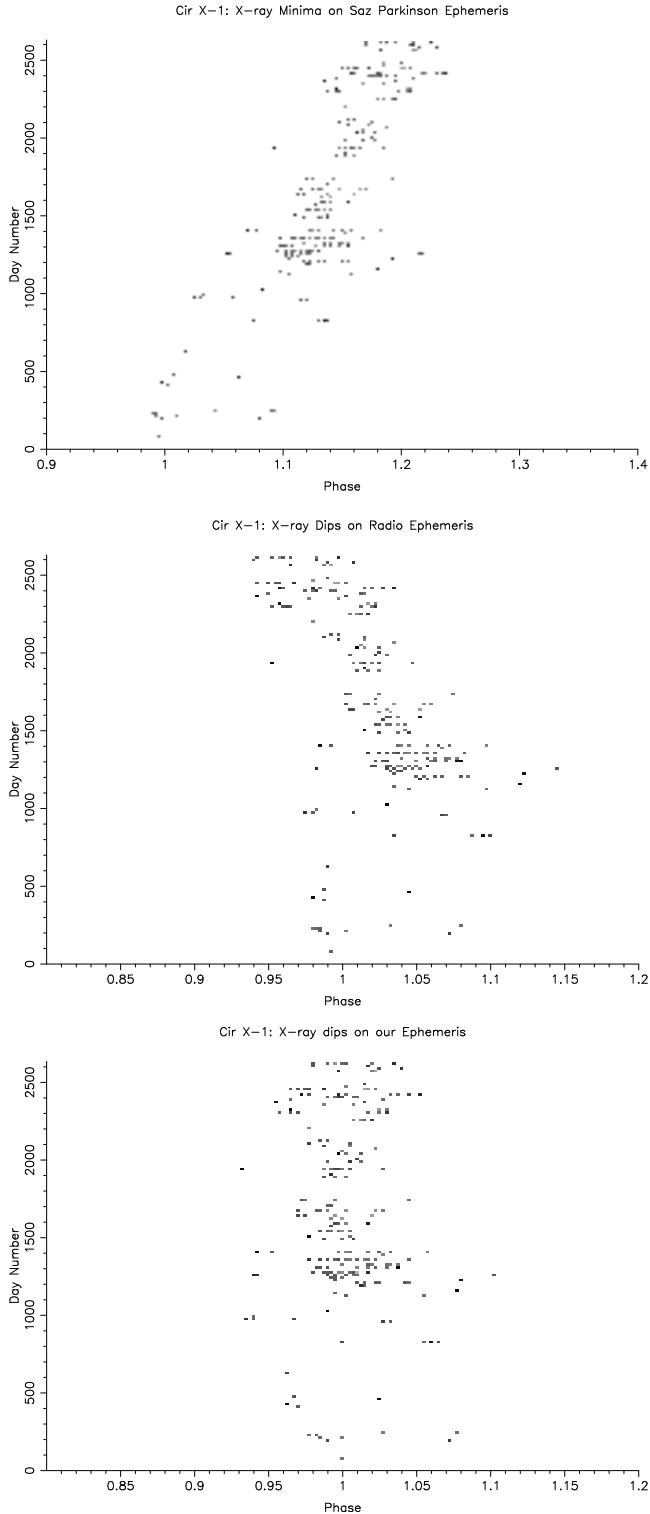

\begin{center}
	\vspace{12pt}
                \psfig{file=fig8a.ps,width=8.5cm,angle=-90}
	\vspace{5pt}
	\vspace{5pt}
                \psfig{file=fig8b.ps,width=8.5cm,angle=-90}
	\vspace{2pt}
	\vspace{5pt}
	     \psfig{file=fig8c.ps,width=8.5cm,angle=-90}
\label{fig:out_dip2}
\caption{{\it Top:} Dynamic lightcurve of X-ray Dips, folded on the SP03 ephemeris. {\it Middle:} X-ray dips folded on the radio ephemeris. {\it Bottom:} X-ray dips folded on our ephemeris. In this figure the grayscale is inversely proportional to the RXTE/ASM count rate.}
\end{center}
\end{figure}



\subsection{Ephemeris based on X-ray Dips}

When folded on the radio ephemeris, we see that the dips show far less
scatter about phase zero than do the maxima (Figures 7 \& 8). Based on
the hypothesis that the dips near phase zero on the radio ephemeris
are in some way related to the binary orbit, and thus should provide a
good system clock, we have attempted to fit an ephemeris on the X-ray
dips alone (as described in section 2.1). The radio flares to some
extent correlate with the X-ray dips (Shirey 1998), suggesting a
causal link, but at present little is known about the mechanisms
producing either the radio flares or X-ray dips, which makes it
impossible to exploit their relative timing. We minimise assumptions
by keeping the time of phase zero as a free parameter in the fit. (If
instead the best fit ephemeris is calculated holding $MJD_0$ at the
same value as the radio ephemeris, the smallest value of the phase
scatter with respect to phase zero is increased by about 15\%). The
result of dip phase minimisation is the following ephemeris:

\begin{equation}
	MJD_N = 50081.76+(16.5732-2.15\times10^{-4}N)N
\end{equation}

\noindent 
with 3$\sigma$ errors on the ephemeris components $MJD_0$, $P$ and $C$
of 8.7 $\times$ 10$^{-2}$d, 3.2 $\times$ 10$^{-3}$d and 2.6 $\times$
10$^{-5}$d respectively (section 2.1).


\begin{figure}
   \psfig{file=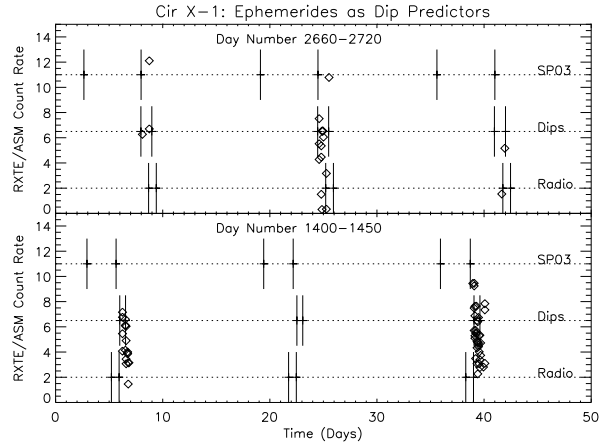,width=8.5cm}
\caption{RXTE/ASM lightcurve at two epochs (diamonds), with 3$\sigma$ bounds on the timing of X-ray dips, as predicted by the three ephemerides (section 4). In each plot the ephemerides, from top to bottom, are based on the X-ray maxima (SP03), X-ray dips (this work) and radio flares (Stewart et al 1991). Ephemeris predictions offset for clarity; no prediction of count rate during dipping is made.}
\end{figure}

We test the predictive power of this X-ray dip ephemeris by plotting
the occurrence of dips against its predictions. As can be seen in
Figures 8 \& 9 , the dip ephemeris predicts the occurrence of dips
with a higher degree of accuracy than the radio or bulk X-ray
ephemerides. In broad agreement with the SP03 ephemeris, we find a
quadratic correction term almost an order of magnitude above that
suggested from the radio ephemeris. It appears that $MJD_0$ for the
X-ray dips occurs approximately 7 hours {\it earlier} than for the
radio flares, but this lag is within the extrapolated error of the
radio ephemeris of $\sim$0.36 days (8.5 hours).

\section{Discussion}

The central issue with Cir X-1 has been the nature of the companion
star, and the ASM lightcurve does not appear immediately to resolve
the issue. Furthermore, the system eccentricity and inclination have
not been constrained (we note that the eccentricity is also only
estimated for A0538-66, the prototypical Be XRB, whose X-ray
lightcurve, when active, is strikingly similar to Cir X-1; Charles et
al 1983). We examine here the use to which the RXTE/ASM lightcurve can
be put to address these issues.

\subsection{The X-ray maxima}

The significant phase wandering of the maxima during the
``transitional'' state of the lightcurve suggests that they are
associated with some feature of the system that is not entirely locked
into the orbital clock. This feature is not completely independent of
the orbital clock either, as there are still {\it no} cycles for which
an X-ray peak appears between orbital phase 0.8 and 0.9, although
during the ``transitional'' interval there are cycles in which the
peak occurs slightly {\it before} phase zero. The X-ray dips appear to
be a better system clock than the outbursts, and we identify their
occurrence with changing mass transfer from the donor during
periastron passage (sections 4.1 \& 5.2). Stellar wind accretion
cannot account for the phase wandering of the maxima; for example, in
the case of GX301-2, an eccentric ($e=0.462$) X-ray binary in which
the accretion takes place via wind plus a small accretion stream
(Leahy 2002), the phase wandering of maxima is tiny compared to that
for Cir X-1. We are thus left with the accretion disk to provide the
source of material for the maxima, as it is capable of doing so in a
manner that could show sufficient variation in the time of occurrence.

The mechanism to do so is not immediately clear. The obvious analogy
would be the dwarf nova cycles seen in cataclysmic variables (Warner
1995), however this interpretation suffers obvious limitations. The
irradiation from the neutron star is likely to keep the inner disk in
a permanent high state, suggesting such cycles would require extremely
fine tuning (e.g. Frank, King \& Raine 2002). Furthermore, this model
does not account for the scaling of the X-ray output at maximum with
the light curve mean (see section 3.1).

We prefer to argue that the X-ray maxima are due intervals of
increased mass transfer onto the compact object as a result of a
combination of disk perturbations and varying mass transfer rate, both
driven to some extent by the eccentric orbit. These variations will
occur at or near the viscous timescale of the disk as this is the
timescale for disturbances to propagate through the disk. This
interpretation has the attraction that it allows the outbursts to
occur over a quasi-steady base level of accretion, as is observed with
RXTE/ASM: such response to perturbations will occur preferentially in
the outer disk, where material is closer to the impact region of the
varying mass transfer stream. This model also has the benefit of
predicting how the amplitude of the maxima will vary with the running
average: both should scale with the global density of matter in the
accretion disk.

It is still unclear what causes the phase of the X-ray maxima to vary
so wildly during the ``transitional'' region, but this interval does
immediately precede the longterm return to low levels of X-ray output,
so therefore may be related to it. However, it is also possible that
the longterm decline is part of an even longer-term pattern of
variation in the mass transfer from the donor, possibly due to tidal
evolution of the system (see section 5.5).

An alternative model for both the longterm variation outburst level
and the phase wandering thus involves Cir X-1 as a Be/X-ray transient,
in which the reformation and destruction of the equatorial disk
provides the mechanism for the variation of the X-ray lightcurve on a
timescale of decades. Variations in the phase of X-ray maxima on a
timescale of years are then produced by the precession of the
equatorial disk and/or evolution of 2D structure in this disk. It is
known that in several Be/X-ray transients with similar orbital periods
to Cir X-1, the equatorial disk is tilted with respect to the plane of
the orbit and precesses (e.g. V0332+53; Negueruela et al 1999, 4U
0115+63/V635 Cas; Negueruela et al 2001, hereafter N01). This disk may
also be destroyed and reformed on a timescale of months - years, as
seen in V635 Cas (N01). Furthermore, 2D structure in equatorial
decretion disks through global one-armed oscillations is predicted
theoretically (Okazaki 2001) and indeed suggested observationally from
V/R variability in several Be/X-ray transients on a timescale of years
to decades (Hanuschik et al 1995). The high and unknown reddening
towards Cir X-1 makes constraining such models extremely difficult for
the forseeable future (c.f. Clark et al 2003).

\subsection{X-ray dips as tracers of $\dot{M}(t)$}

Using the X-ray dip ephemeris, the dip occurrence rate depends
strongly on orbital phase, as all dips are clustered around $\pm
\sim$1 day of phase zero. That this is in sharp contrast to the maxima
used by SP03 to determine the ephemeris does not invalidate their
discussion, as we are examining a different feature of the
lightcurve. We examine here the information that can be extracted
about the binary components from this property.

Absorption dips have been used as a diagnostic in the RXTE/ASM
lightcurve of the near-circular orbit black hole candidate Cyg X-1
(Balucinska-Church et al 2000). In this case the occurence rate of
X-ray dips was also shown to vary strongly with orbital phase, in a
manner dependent on the line of sight to the binary: the dips occured
with greatest frequency when the donor passed directly in front of the
accretor. In this model the X-ray dips are due to photoionisation in
the wind of the donor. In the case of Cir X-1, however, the
near-coincidence of the dips with times of maximum activity at other
wavelengths (and thus periastron passage) suggests the semimajor axis
of the binary orbit would have to be almost aligned with our line of
sight to the binary, which we reject as requiring fine tuning.

However, subsequent studies of Cyg X-1 with RXTE/PCA divide the X-ray
dips in Cyg X-1 into two populations based on spectral behaviour:
``Type A'' dips, concentrated about donor interposition phases as
above, and ``Type B,'' which occur throughout the orbit (Feng \& Cui
2002). One suggestion put forward for these dips is absorption of an
X-ray emitting region by optically thick ``clouds'' in the accretion
flow near the accretion disk. The number density of such clouds in the
vicinity of accretion stream impact in the outer disk scales strongly
with $\dot{M}$ (equation 10 of Frank, King \& Lasota 1987, hereafter
FKL87). In an eccentric system such as Cir X-1, $\dot{M}$ will be a
strong function of orbital phase (see Charles et al 1983 or Brown \&
Boyle 1984). We thus identify the X-ray dips exhibited by Cir X-1 as
``Type B'' dips, which here occur only during episodes of enhanced
mass transfer at periastron passage. This has the attraction of
removing the need for fine-tuning of the line of sight to the system.

The accretion disk will respond to changes in $\dot{M}$ on its viscous
timescale. We can estimate this value from the phase spread in the
outbursts (Figure 7 and section 5.1) or alternatively from assumptions
on the mass transfer rate and viscosity (Frank, King \& Raine 2002),
to be of order 10 days. The inner disk can respond more quickly but is
likely to be highly ionised (Brandt \& Schulz 2000), thus will not
figure in the formation of the X-ray dips. The response of the
accretion disk to varying $\dot{M}$ thus occurs on too long a
timescale to affect the relationship between $\dot{M}(t)$ and the
occurrence rate of the dips.

The velocity of the gas flow in the accretion stream will be greater
than the adiabatic or isothermal sound speeds, allowing internal
shocks to be neglected in the direction of the flow (c.f. Frank, King
\& Raine 2002; we assume further that any variation in sound speed
along the stream due to changing $\dot{M}$ is insufficient to bring
the sound speed above the freefall velocity at any point in the
stream). The sound travel time across the stream will be long compared
to the transfer time to the accretion disk (c.f. Lubow
\& Shu 1975), so we neglect the internal response of the gas in the
accretion stream. The impact of the accretion stream on the disk
results in shock formation and the condensation of cold absorbing
clouds on a timescale similar to the recombination timescale $t_{rec}$
(FKL87), $\sim$ 1-30 minutes for the parameters of Cir X-1. The
evaporation timescale for such clouds is of order a few dynamical
timescales $t_{dyn}$ (FKL87, Krolik, McKee \& Tarter 1981), so they
will persist for $\ga$5 hours. The formation rate of clouds leading to
X-ray dips will thus respond quickly to changes in $\dot{M}$, and
furthermore such dips will persist for a long enough interval to
affect the X-ray lightcurve. The $\ga$ 5 hour timescale for cloud
evaporation suggests reduction in dip occurrence rate may lag shortly
behind the reduction in $\dot{M}$, which is fully consistent with the
slight apparent asymmetry in the dip population (Figures 8 \& 10).

\subsection{Using the dips to investigate the binary parameters}

We attempt in this section to relate the rate of occurrence of X-ray
dips to the donor mass $M_2$, the eccentricity $e$ and the donor
temperature $T$. To do so we require a model predicting $\dot{M}(t)$
as a function of these parameters. A preliminary theoretical
investigation of $\dot{M}(t)$ due to tidal lobe overflow in an
eccentric-orbit binary was undertaken by Brown \& Boyle (1984;
hereafter BB84). They model the mass transfer as as overflow across an
instantaneous tidal lobe (analogous to the Roche Lobe in the circular
case), with the constraint added that matter must move with positive
velocity relative to the instantaneous tidal lobe, for accretion to
occur. This estimates the effect of the changing size of the tidal
lobe throughout the binary orbit. This correction also has the effect
of shifting the peak in $\dot{M}$ slightly earlier than true
periastron. This semianalytic approach was followed up by numerical
simulation (Boyle \& Walker 1986), exploring the range of parameter
space for which capture of transferred matter might occur, in an
attempt to model the X-ray activity of A0538-66. Only a single binary
orbit was followed in these simulations, so it is not possible to
determine the fate of material not immediately captured by the neutron
star. Earlier simulations (Haynes, Lerche \& Wright 1980) did not take
into account hydrodynamic effects, but in covering more orbits
suggested that matter not immediately accreted in one periastron
passage may be accreted in subsequent cycles, complicating the
accretion profile. We thus reject the capture condition espoused by
Boyle \& Walker (1986), in which gas is only captured if its specific
potential energy is greater in magnitude than its specific kinetic
energy. More modern simulations of eccentric orbit accreting systems
suggest this subsequent accretion may be extremely important in
determining the X-ray output of an eccentric binary (Podsiadlowski
2003 priv. comm.). However, as our interest is not in the X-ray
generation rate per se but in absorption due to the matter stream, we
take the BB84 model for lack of current alternatives. The reader is
therefore cautioned that the model on which our constraints will be
based is in some sense a preliminary one.

\subsection{A simple model}

We attempt to fit the profile of dip occurrence rate as a function of
$M_2$ and $e$. The major source of uncertainty in the BB84 model is
the exponential dependence of $\dot{M}(t)$ on the factor

\begin{equation}
	\beta = \frac{a(1-e)}{H_p}
\end{equation}

\noindent where $H_p$ is the (as yet unknown) pressure scale height 
at the surface of the donor at periastron passage ($H=kT/m_{p}g$) and
$a$ the semimajor axis of the orbit. We make the assumption here that
the donor fills its tidal lobe at periastron, allowing us to use the
Roche Lobe prescription (Eggleton 1983) for the radius of the
donor. The known orbital period thus allows us to place the following
constraint on the factor $\beta$:

\begin{equation}
	\beta \simeq 106\left(\frac{q}{f^2(q)(1+q)^{1/3}}\right)
	\left(\frac{1}{T_4(1-e)}\right)
\end{equation}

\noindent where $f(q)$ is the tidal lobe radius at periastron as a fraction of the
semimajor axis (Eggleton 1983), $q=M_2/M_1$ and $T_4$ is the
temperature at the surface of the donor, at periastron, in units of
10$^4$ K. As the X-ray dips occur over a very limited phase range, we
neglect any radius variation that may occur in the donor over the
course of the periastron passage, beyond the variation of the tidal
lobe itself. Furthermore, we follow Charles et al (1983) in assuming
the surface of the donor corotates with the orbit. It is currently
unclear what the quantitative effect of non-synchronous donor rotation
will be on the $\dot{M}(t)$ profile. We expect the change in the
amount of matter ejected from the surface of the donor to compete with
an opposing change in the amount of matter that can be captured by the
accretor (e.g. Petterson 1978), however the balance of the effects is
far from obvious. In the absence of a model for the phase dependence
of this behaviour, we retain the assumption of corotation. We merely
caution here that if Cir X-1 is a dynamically young system, as
suggested by its apparent eccentricity and other behaviours (section
5.5), non-synchronous rotation is a distinct possibility.

We plot the dip occurrence profile, along with an example $\dot{M}(t)$
prediction, in Figure 10. As can be seen, the profile can be fit quite
readily with the tidal lobe model, within the errors (assumed
gaussian) between the measured X-ray dip rate and the intrinsic cloud
formation rate (section 5.2). For the following analysis, dips
occurring in the phase range (0.95 - 1.05) are used, as outside this
region the errors are comparable to the observed dip rate. Clearly the
fit is degenerate, in that in principle any combination of $q,T_4$ and
$e$ can reproduce the observed dip profile. We narrow the parameter
space somewhat by noting that the donor mass is unlikely to exceed the
$\sim 15 M_{\odot}$ necessary to form a neutron star (Tauris \& van
den Heuvel 2003) or else it would have noticeably evolved. More
stringent constraints on the donor mass through direct detection of
the sytem are not possible due to the uncertain location of the
IR-emitting regions (Clark et al 2003) and the uncertain level of
reddening (Glass 1994).

\begin{figure}
   \psfig{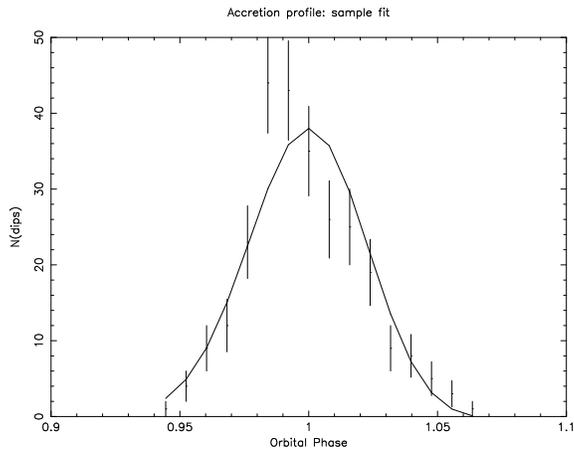}
\caption{X-ray dip occurrence profile as a function of (presumed orbital) phase. Overplotted is a sample prediction of the BB84 model, in this case $M_2$=2, e=0.4, $H_p$=7.5 $\times$ 10$^4$ km.}
\end{figure}


A large range of $T_4$ is in principle possible, as there are several
sources of energy dissipation that could heat the surface of the donor
significantly above its unperturbed value of $T_4 \sim 0.5 - 1$. The
accretor has been among the brightest X-ray sources in the sky, so
might be expected to irradiate the donor to $T_4 \ga$ 10 (e.g. van
Teeseling \& King 1998; we neglect here the enhanced mass loss such a
mechanism might produce). The supernova itself may have deposited a
large amount of energy in the donor leading to significant heating of
the stellar envelope to $T_4 \sim$10 (Marietta et al 2000,
Podsiadlowski 2003). Finally, if the system really is eccentric, tidal
circularisation will give rise to internal heating of the donor, also
bringing $T_4$ to $\sim$ 10 or higher (see Podsiadlowski 1996). To
account for these possibilities, we repeat the $\dot{M}(t)$ profiles
for several input values of $T_4$ in the range (0.1 $\le$ $T_4$ $\le$
100). For each combination of parameters, the least-squares fit to the
dip profile is evaluated and the parameters that fit the profile most
closely retained.

The results are plotted in Figure 11. We note that the condition of
tidal lobe overflow at periastron leads to larger scale heights than
is usually the case ($\sim$ 10$^4$ km), though still small compared to
the stellar radius. The reader is urged to bear in mind that the low
number of dips recorded per bin by RXTE/ASM leads to significant
uncertainty in the true dip occurrence profile (Figure 10). Monte
Carlo simulations suggest that the values of $M_2(e)$ calculated from
fitting these dips are thus uncertain to a 3$\sigma$ error of at least
50\%. Above $T_4 \sim$5 the factor $\beta$ becomes largely insensitive
to the donor mass $M_2$, and {\it all} combinations occur at $e
\ga$ 0.6. We note further that the values of $M_2(e)$ should be
considered lower limits, as the true $\dot{M}$ profile may be narrower
than the X-ray dip occurrence profile (see section 5.2), which would
shift the curves upwards in $M_2$. We thus make no attempt to
determine the value of $M_2$ from our simulations, opting instead to
determine regions of $e$ for which solutions exist for reasonable
values of $T_4$.

\begin{figure}
   \psfig{file=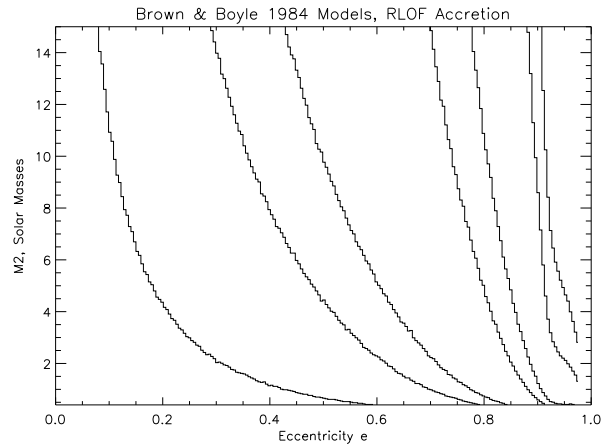,width=8.5cm}
\caption{Combinations of orbital eccentricity and donor mass that fit the observed dip occurrence rate profile. From left: $T_4$=0.1, 0.5, 1.0, 5.0, 10.0, 50.0 and 100.0 respectively.}
\end{figure}

As expected, for the same width of $\dot{M}$ profile, larger
eccentricities correspond to lower values of $M_2$ and thus lower
donor radius. Perhaps surprisingly, low eccentricities are not in
principle ruled out, although they do require both low surface
temperatures and large donor masses. However, the surface of the donor
is likely to be significantly heated by any mechanisms discussed above
(so $T_4 \ga$ 5). Thus, within the limitations of the model (see
above), the system is likely to be at high eccentricity (0.6 $\la e
\la 0.95$) for all $M_2$ (Figure 11). More accurate limits await the
results of more detailed simulations of $\dot{M}(t)$ for highly
eccentric x-ray binaries.

\subsection{Is Cir X-1 a very young system?}

The results from the above toy model suggest the system must be in a
state of high eccentricity, leading to a natural interpretation with
Cir X-1 in an early stage of post-supernova evolution. We urge the
reader to bear in mind the limitations of the above model, however; in
particular the assumption of corotation will likely not hold for a
young system. In the case of a non-irradiated donor with a thick
convective envelope, the synchronisation timescale can be estimated at
$\ga 10^6$ years for a 16-6 day orbit (Zahn 1977), suggesting the
effects of nonsynchronous donor rotation should be considered before
general limits can be placed on the binary orbit. In particular the
balance between the competing effects of non-synchronous rotation will
likely depend on orbital phase. Without a robust exploration of the
effect of non-synchronous donor rotation on the accretion profile it
is impossible to place limits on the discrepancy between our
simulations and the true accretion profile. Our limits on the
eccentricity thus apply only to the special case in which corotation
has been achieved but circularisation has not. If the system were much
older than $\sim 10^6$ years, this case might apply: the
synchronisation timescale for a 16.6-day orbit is a factor $\sim
10^{2-4}$ longer than the circularisation timescale for the range of
$q$ used in the above model (Zahn 1977).

The ephemeris based on the X-ray dips is a stronger indicator of the
age of Cir X-1. The X-ray dip ephemeris establishes a characteristic
timescale $P/2\dot{P}$ even shorter than that suggested by the SP03
analysis, at $\sim$1000 years. If the system were in a state of tidal
evolution shortly after the supernova, the system components may not
yet have achieved corotation and alignment of their spin axes. The
evolution to alignment is expected to give rise to high amplitude
variations in mass transfer on a timescale of decades or more
(Podsiadlowski 2003, priv. comm.). A look at the long-term X-ray
history of Cir X-1 (e.g SP03) shows variability by a factor of at
least 10 on such a timescale. This may be further evidence that the
system is evolving dynamically, though without ruling out other
explanations for such long-term apparent variation, no definite claim
can be made.

However, Cir X-1 is surrounded by a synchrotron radio nebula
approximately 5 $\times$ 10 arcmin$^{2}$ (Stewart et al 1993). Now
that the association with SNR G321.9-0.3 has been ruled out (Mignani
et al 2002), we propose that this radio nebula may in fact be the
supernova remnant of Cir X-1. (When completing this paper we became
aware (Fender 2003, priv. comm.) of others exploring this possibility,
but believe it has not yet been discussed in the literature.) Such a
conclusion is consistent with standard SNR expansion models, and the
high reddening would obscure optical emission from the nebula. With a
diameter of $\sim$ 8 pc, the remnant would likely be in the
Sedov-Taylor stage, which implies $\ga$ 4,000 years have elapsed since
the supernova (e.g. Choudhuri 1998). Comparison with observed SNRs at
this stage of evolution (e.g. Hughes, Hayashi \& Koyama 1998) suggest
the X-ray luminosity is likely to be $L_X \sim 1-50 \times 10^{35}$
erg s$^{-1}$, or roughly 1\% the ``steady'' output of the point
source. A study of archival XMM and Chandra observations of Cir X-1
(Clarkson \& Charles 2003) shows that, in the region occupied by the
radio nebula, X-ray emission is dominated by ISM scattering of the
bright central point source. This scattered component is bright enough
to mask any X-ray SNR at the location of Cir X-1, allowing no direct
determination of the existence or otherwise of an X-ray counterpart to
the nebula. The light travel time of reflected paths is long enough to
cause pessimism of the chances of sucess of a search for a SNR during
intervals of dipping.

\subsection{The External Magnetic Field of the Neutron Star}

The primary objection to the young source scenario has always been
that the external magnetic field of the neutron star is observed to be
much weaker than expected for a young system, as suggested by the
existence of Type I bursts (Tennant et al 1987) and Z-source QPOs
(Shirey, Bradt \& Levine 1999), and the repeated non-detection of
X-ray pulsations (though the latter is inclination-dependent). We
suggest the discrepancy may be resolved by accretion-induced screening
of the core magnetic field. Detailed consideration of the accretion
and screening process is beyond the scope of this work: we merely
summarise the model and recent results. Matter initially channeled
onto the magnetic poles spreads under the weight of the accretion
column to cover the surface of the neutron star. This matter is a
fully ionised plasma, and therefore is diamagnetic, leading to the
screening of the internal magnetic field (Taam \& van den Heuvel 1986:
see Choudhuri \& Konar 2002 for a review). Advection of the screening
layer then submerges the internal magnetic field (Romani
1990). Evolution of the external field then depends on the competition
between the timescales for advection of the accreted matter and for
ohmic diffusion of the internal field into the newly accreted matter
(Konar \& Bhattacharya 1997). The internal field will break through
the screening on the timescale for ohmic diffusion across the outer
crust (e.g. Cumming, Zwiebel \& Bildsten 2001), which will be measured
in decades for all sensible values of the crustal scale height. Brown
\& Bildsten (1998) compute the competition between advection and ohmic
diffusion for polar accretion, and find that for $\sim
0.5\dot{M}_{Edd}$ the timescale for ohmic diffusion is larger than for
advection, suggesting screening may be possible on the advection
timescale ($\sim 10^3-10^5 yr$). Recent 2D simulations (Choudhuri \&
Konar 2002) suggest a timescale of $10^{5}$ years for screening by 3-4
orders of magnitude. Further work is needed to arrive at robust
timescales for screening: in particular the evolution of instabilities
breaking the screening (c.f. Cumming, Zweibel \& Bildsten 2001) and
the extension of the model to three dimensions have yet to be
explored. Without information about the likely mass transfer rate of
Cir X-1 immediately after the supernova, it is difficult to judge the
likely effectiveness of screening, as no information is available as
to the $\dot{M}$ history before the age of X-ray astronomy. The
RXTE/ASM count rate and source distance suggest instantaneous
accretion rates approach $\sim3 \dot{M}_{Edd}$ in the early stages of
the RXTE/ASM lightcurve, while the ``steady'' accretion rate appears
to reach $\sim 0.7 \dot{M}_{Edd}$ during the ``high'' state (section
3). If this is an example of repeating behaviour (section 5.5), then
screening of the internal magnetic field by several orders of
magnitude appears at present a possibility, on a timescale ranging
perhaps from 10$^3$ - 10$^5$ years. Furthermore, we remind the reader
that during enhanced accretion at periastron passage, the mass
transfer onto the accretor may be spherical even with strong internal
B-fields. We speculate that should screening be established, this
effect would further counteract outward diffusion of the internal
field, as screening matter would no longer go through the stage of
spreading from the poles.

\section{Conclusion}

We have examined the periodic nature of the X-ray maxima and dips
exhibited in the RXTE/ASM lightcurve of Circinus X-1. We find that the
X-ray dips provide a more accurate system clock than the maxima, and
thus identify them with the periastron approach of the binary
components. The outbursts are interpreted as the viscous timescale
response of the disk to perturbations from the varying mass
transfer. We attempt to use the X-ray dips to constrain the nature of
the donor and the eccentricity of the orbit, and find that, within the
limitations of the model used, a wider range of parameters are in
principle allowed than has previously been thought, although without
further exploration of the effects of nonsynchronous donor rotation,
quantitative limits are at this stage unreliable. The high rate of
change of the orbital period adds to the growing body of evidence that
Cir X-1 is in an extremely early stage of post-supernova evolution. If
the system is indeed in an early stage of post-SN evolution, the
observed weak external magnetic field must suggest strong screening of
the internal magnetic field of the neutron star.

\section{Acknowledgements}

WIC thanks Philipp Podsiadlowski and Kinwah Wu for exciting and
informative discussions on eccentric-orbit binaries, and Katherine
Blundell for discussion on the radio appearance of young supernova
remnants. The authors acknowledge Malcolm Coe for insightful comments
during the early stages of this work. We also thank the referee, Peter
Jonker, for useful comments and suggestions. WIC and NO acknowledge
the support of PPARC studentships.

\label{lastpage}


\begin{thebibliography}{}

\bibitem{}
Balucinska-Church M., Church M.J., Charles P.A., Nagase F., LaSala J,. Barnard R., 2000 MNRAS 311, 861 
\bibitem{}
Boyle C.B., Walker I.W., 1986 MNRAS 222, 559 
\bibitem{}
Brandt W.N., Schulz N.S., 2000 ApJ Lett 544, 123 
\bibitem{}
Brandt W. N., Fabian A. C., Dotani T., Nagase F., Inoue, H., Kotani
T., Segawa Y., 1996 MNRAS 283, 1071
\bibitem{}
Brown J.C., Boyle C.B., 1984 A\&A 141, 369 (BB84) 
\bibitem{}
Brown, E.F., Bildsten L., 1998 ApJ 496, 915
\bibitem{}
Choudhuri A. R., Konar S., 2002 MNRAS 332, 933
\bibitem{}
Choudhuri A.R., {\it The Physics of Fluids and Plasmas}, 1st edition,
1998 CUP
\bibitem{}
Charles, P.A. et al, 1983 MNRAS 202, 657
\bibitem{}
Clark J.S., Charles P.A., Clarkson W.I., Coe M.J., 2003 A\&A 400, 655 
\bibitem{}
Clarkson W.I., Charles P.A., Laycock S., Coe M.J., Tout M.T., Wilson, C., 2003a MNRAS 339, 447 
\bibitem{}
Clarkson W.I., Charles P.A., Laycock S., Coe M.J.,  2003b MNRAS 343, 121
\bibitem{}
Clarkson W.I. \& Charles P.A., 2003 MNRAS, in prep
\bibitem{}
Cumming A., Zweibel E., Bildsten L., 2001 ApJ 557, 958
\bibitem{}
Eggleton P.P., 1983 ApJ 268, 368
\bibitem{}
Fender R.P., Spencer R.,Tzioumis T., Wu K., van der Klis M., van Paradijs J., Johnston H., 1998 ApJ Lett, 506, 121 
\bibitem{}
Fender, R.P., 1998 IAU Colloquium 164, eds. Zensus J.A., Taylor G.B., Wrobel J.M., astro-ph/9707317
\bibitem{}
Feng Y.X., Cui W., 2002 ApJ 564, 953 
\bibitem{}
Frank J., King A.R., Raine, D.J., {\it Accretion Power in Astrophysics}, 3rd edition, 2002 CUP
\bibitem{}
Frank J., King A.R., Lasota J-P., 1987 A\&A 178, 137 (FKL87)
\bibitem{} 
Glass I.S., 1994 MNRAS 268, 742 
\bibitem{}
Goss W.M., Mebold U., 1977 MNRAS 181, 255
\bibitem{}
Hanuschik R. W., Hummel W., Dietle O., Sutorius E., 1995 A\&A 300, 163
\bibitem{}
Haynes R.F., Lerche I., Wright A.E., 1980 A\&A 81, 83 
\bibitem{}
Haynes R.F., Jauncey D.L., Murdin P.G., Goss W.M., Longmore A.J., Simons L.W.J., Milne D.K., Skellern, D.J., 1978 MNRAS 185, 661
\bibitem{}
Hughes John P., Hayashi I., Koyama, K., 1998 ApJ 505, 732
\bibitem{}
Iaria R., Burderi L., Di Salvo T., La Barbera A., Robba N.R., 2001 ApJ 547, 412
\bibitem{}
Johnston H.N., Fender R.P., Wu, K., 1999 MNRAS 308, 415
\bibitem{}
Kaluzienski L.J., Holt S.S., Boldt E.A., Serlemitsos P.J., 1976 ApJ Lett. 208, 71 
\bibitem{}
van der Klis, M., 1995, in {\it X-ray Binaries}, eds. Lewin W.H.G., van Paradijs J., van den Heuvel P.J., first edition, CUP, p252
\bibitem{}
Konar, S., Bhattacharya D., 1997 MNRAS 284, 311
\bibitem{}
Krolik J. H., McKee C. F., Tarter C. B., 1981 ApJ 249, 422
\bibitem{}
Kuulkers, E., Wijnands, R., Belloni T., Mendez M., van der Klis M.,
van Paradijs J., 1998 ApJ 494, 753
\bibitem{}
Leahy D.A., 2002 A\&A Lett 391, 219
\bibitem{}
Levine, A. M., Bradt, H. V., Enevoldsen, A., Morgan, E. H., Remillard,
R. A., Wen, L., Smith, D. A., 2000 HEAD \#32, \#43.07
\bibitem{}
Levine A. M., Bradt H., Cui W., Jernigan J. G., Morgan E. H.,.; Remillard R., Shirey R. E., Smith, D. A., 1996 ApJ Lett 469, 133
\bibitem{}
Lubow S.H \& Shu F.H., 1975 ApJ 198, 383
\bibitem{}
Margon B., Lampton M., Bowyer S., Cruddace R., 1971 ApJ Lett 169, 23
\bibitem{}
Marietta E., Burrows A., Fryxell B., 2000 ApJS 128, 615 
\bibitem{}
Mirabel I.F., Rodriguez L.F., 1999 ARA\&A 37, 409 
\bibitem{}
Mignani R.P., De Luca A., Caraveo P.A., Mirabel I.F., 2002 A\&A 386, 487 
\bibitem{}
Moneti A., 1992 A\&A Lett. 260, 7 
\bibitem{}
Negueruela I., Okazaki A. T., Fabregat J., Coe M. J., Munari U., Tomov T., 2001 A\&A 369, 117 (N01)
\bibitem{}
Negueruela I., Roche P., Fabregat J., Coe M. J., 1999 MNRAS 307, 695
\bibitem{}
Okazaki, A., 2001 PASJ 53, 119
\bibitem{}
Petterson J., 1978 ApJ 224, 625
\bibitem{}
Podsiadlowski Ph., 2003 MNRAS in press, astro-ph/0303660
\bibitem{}
Podsiadlowski. Ph., 1996 MNRAS 279, 1104 
\bibitem{}
Romani R.W., 1990 Nature 347, 741 
\bibitem{}
Saz Parkinson P.M et al., 2003 astro-ph/0303402 (SP03)
\bibitem{}
Scargle J.D., 1989 ApJ 343, 874 
\bibitem{}
Scargle J.D., 1982 ApJ 263, 835 
\bibitem{}
Schulz N.S., Brandt W.N., 2002 ApJ 572, 971
\bibitem{}
Shirey R.E., Bradt, H.V., Levine, A.M., 1999 ApJ 517, 472 
\bibitem{} 
Shirey R.E., Levine, A.M., Bradt H.V., 1999 ApJ 524, 1048 
\bibitem{}
Shirey R.E., Bradt, H.V., Levine, A.M., Morgan, E.H., 1998 ApJ 506, 374 
\bibitem{}
Shirey R.E., 1998 PhD Thesis, MIT 
\bibitem{}
Stewart R.T., Caswell J.L., Haynes R.F., Nelson G.J., 1993 MNRAS 261, 593 
\bibitem{}
Stewart R.T., Nelson G.J., Pennix W., Kitamoto S., Miyamoto S., Nicolson G.D., 1991 MNRAS 253, 212
\bibitem{}
Taam R. E., van de Heuvel E. P. J., 1986 ApJ 305, 235
\bibitem{}
Tanaka, Y., Lewin, W.H.G., 1995, in {\it X-ray Binaries}, eds. Lewin W.H.G., van Paradijs J., van den Heuvel P.J., first edition, CUP, p126
\bibitem{}
Tauris \& van den Heuvel 2003, astro-ph/0303456, to appear in {\it Compact Stellar X-ray Sources}, eds. Lewin W.H.G \& van der Klis, M., CUP
\bibitem{}
Tauris T.M., Fender R.P., van den Heuvel E.P.H., Johnston H.M., Wu K.,
MNRAS 1999, 310, 1165 
\bibitem{}
van Teeseling A. \& King A.R., 1998 A\&A 338, 957 
\bibitem{}
Tennant A.F., 1987 MNRAS 226, 971 
\bibitem{}
Tennant A.F., Fabian A.C. \& Shafer R.A., 1986a MNRAS 221P, 27 
\bibitem{}
Tennant A.F., Fabian A.C. \& Shafer R.A., 1986b MNRAS 219, 87 
\bibitem{}
Warner 1995, {\it Cataclysmic Variables}, first edition, 1995, CUP 
\bibitem{}
Whelan 1977, J.A.J. et al., MNRAS 181, 259
\bibitem{}
Zahn, J-P., 1977 A\&A 57, 383

\end{thebibliography}
\end{document}